# Artificial Intelligence and the New Science of Culture

Douglas R. Guilbeault[1], Bhargav Srinivasa Desikan[2], James A. Evans[3]

[1] *Graduate School of Business, Stanford University, U.S.*

[2] *Oxford Internet Institute, University of Oxford, U.K.*

[3] *Sociology & Data Science, University of Chicago, U.S.*

Correspondence: dguilb@stanford.edu, bhargav.srinivasadesikan@oii.ox.ac.uk, jevans@uchicago.edu

## Abstract

Cognition and culture have long been treated as parallel objects of study, joined more by metaphor than by mechanism. We argue they are co-constituted in a manner now empirically tractable: subjectivities can be measured as local trajectories through a high-dimensional cultural field, and the field is itself the aggregate of those trajectories. Recent advances in machine learning, including embedding methods and large generative models, provide the first general framework for measuring this co-constitution from micro-cognition to macro-social structure. Vector geometry recovers individual conceptual structure, organizational communication, and large-scale ideologies; captures multimodal cultural content beyond text; and generates testable predictions about cultural emergence, including the simultaneous arrival of independent discoveries across distant minds. We treat this predictability of simultaneous innovation as evidence for the co-constitutive view: when the geometry of the cultural field is measurable, trajectories of cognitive search through it become predictable. We then examine generative AI agents as simulators of human subjectivity and as a qualitatively new coordinating substrate whose insertion into social life introduces evolutionary dynamics unprecedented in human cultural history. We formalize this reflexive condition as a Heisenberg-like Uncertainty Principle for generative AI: as instruments for fixing a culture's position grow more precise, our capacity to predict its trajectory degrades, because the machinery of cultural description and cultural action have merged. We close with ethical questions raised by AI-mediated cultural drift, recursive synthetic culture, and the limits of human cultural agency, offering guidelines for safe, equitable, and privacy-preserving research on AI and culture.

**Keywords:** artificial intelligence, cultural evolution, culture and cognition, generative AI, large language models, simultaneous discovery, social prediction, word embeddings

## Introduction

Interest in the study of culture has grown sharply across sociology, psychology, anthropology, computer science, biology, and management, yet the concept of culture has remained notoriously difficult to pin down. Initially referring to the way of life of specific peoples, culture has come to encompass values, practices, norms, symbols, and tools that constitute human social life and subjectivity (Geertz 1973; DiMaggio 1997; Patterson 2014; Sewell 2005). Some scholars view culture as so broad as to be the sum total of collective human experience (Cronk 2019). Others have judged it impossible to measure or theoretically incoherent (Emirbayer 1997; Vaisey 2021; Adkisson 2014). Parallel impasses persist within cognitive science, where researchers have asked whether anthropology earned its seat at the table (Beller, Bender, and Medin 2012), and within sociology's "culture and cognition" subfield, where critics contend that decades of work have failed to deliver a coherent empirical program (Lizardo 2014; Vaisey 2021).

We revisit these concerns in light of recent advances in machine learning and computational social science. Our central claim is theoretical: cognition and culture are co-constituted in a measurable way. Subjectivities are local trajectories through a high-dimensional cultural field, and the field is itself the aggregate of those trajectories. This claim was previously a programmatic commitment whose implications could not be tested at scale. We argue it is now empirically tractable due to recent advances in artificial intelligence, and its predictions are being borne out.

Clear evidence comes from a striking class of recent findings. When the geometry of the cultural field around a scientific community is measured from its publications, it becomes possible to predict not only which papers will be impactful (Belikov, Rzhetsky, and Evans 2022; Shi and Evans 2023; Sourati and Evans 2023) but also which papers will appear as twins, nearly simultaneous discoveries by independent minds occupying overlapping neighborhoods of the same epistemic terrain (Ogburn & Thomas 1922; Merton 1961, 1963; Kim, Kojaku, and Ahn 2026). When the categorical preferences of partisan agents are simulated using a language model trained on a prior cultural moment, it is possible to forecast the arguments and attitudes those partisans will produce in subsequent crises (Kozlowski, Kwon, and Evans 2024). When categories are invented from scratch by large groups of disconnected individuals, the resulting category systems converge with surprising regularity, even when individuals' own preferences and cognitive associations vary considerably (Guilbeault, Baronchelli, and Centola 2021; Guilbeault & Centola 2025). These results are evidence for a unifying claim: cognitive search is constrained by the cultural field through which it moves, and the geometry of this field and its impacts are now measurable.

We develop the implications of this claim throughout this article. We review embedding methods and large generative models as the technical apparatuses that have made this computational framework empirically operative, drawing the connection to a longer cognitive science tradition that represents concepts as positions in geometric space (Gärdenfors 2000, 2014; Murphy 2002; Piantadosi et al. 2024; Griffiths 2026). We show that a common mathematical framework applies across micro-cognitive, meso-organizational, and macro-cultural scales, and across sensory (e.g., visual, auditory, olfactory), communicative (e.g., text, images), and behavioral modalities[1]. We discuss the constraints that embodied cognition places on what current AI systems can capture, and we engage with the debate over whether and how distributional learning can ground meaning (Bender and Koller 2020; Pavlick 2023; Mitchell and Krakauer 2023; Kadambi et al. 2026). We treat the predictability of simultaneous discoveries as a quantifiable formulation of the historian Paul Forman's (1971) argument that ambient culture shapes the imaginability of scientific ideas and constrains the progress of knowledge. And we examine generative AI agents less as humanlike interlocutors and more as a new and partly alien coordinating substrate whose insertion into human social life is likely to introduce qualitatively novel constraints and evolutionary dynamics into culture itself (Brinkmann et al. 2023; Farrell, Gopnik, Shalizi, and Evans 2025).

Running through the article is a distinction between AI as a mirror and as an engine of culture (MacKenzie 2006). As mirrors, predictive machine-learning systems have become the workhorse of data-driven science, but as engines, the generative interactive chat and now agentic systems built atop them have become the public face of 'AI.'[2] These are not two technologies but the same machinery read in opposite directions, and together, they yield what we describe below as a

1 We note that we do not intend to claim that the current mathematical framework of vector-based embeddings is unified across AI models and their domain applications, nor that such techniques are in their final stage of development. Future innovations are expected, and different geometrical approaches (e.g., hyperbolic or non-Reimannian methods) may prove to be better representational formats for extracting patterns in some domains, e.g., visual experience (Bujack et al. 2022; Nadler, Darragh-Ford, Srinivasa Desikan et al. 2023), social and semantic networks (Sala, De Sa et al. 2018; Chami, Ying et al., 2019), and hierarchical technological systems (Cao and Evans 2026), but not others. Our emphasis, rather, is that the widespread adoption of vector-based embeddings has provided a common, spatial computational language for encoding patterns across a rich variety of modalities and forms, enabling unprecedented convergence in the formal, methodological techniques for extracting meaningful relationships across fields, from individual-level cognitive science to sociological research on large-scale human collectives.

2 We borrow the mirror and engine dichotomy from MacKenzie (2006) who characterized financial algorithms in the same way.

Heisenberg-like Uncertainty Principle for AI: the more precisely our models measure and fix the position of a culture, the less reliably they can forecast its trajectory, because the instruments of understanding and prediction become embedded, endogenous participants in the culture itself. We close with ethical questions raised by recursive feedback loops between humans and machines, and offer guidelines for safe, equitable, and privacy-preserving research at the AI–culture interface.

We make two distinct theoretical commitments throughout. The first is that LLM-based simulations of human cognition are valuable not insofar as they replicate human micro-cognition step by step, an empirical question on which the evidence is mixed (Shiffrin and Mitchell 2023; Bowers et al. 2022; Wang, Morgenstern, and Dickerson 2025; Guilbeault, Caplan, and Yang 2026), but insofar as they measure the *aggregate* statistical structure that shapes macro-cultural dynamics (Piao et al. 2026). The second is that AI systems are not best modeled as humanlike minds, but as a distinct cultural and social technology with their own emergent dynamics (Farrell et al. 2025; Brinkmann et al. 2023). These two commitments are in mild tension, and we hold them in tension deliberately: the first justifies AI as a predictive instrument for cultural science; the second justifies treating AI's introduction into human life as an evolutionary event whose consequences cannot be inferred from how well it imitates us. The tension is not incidental. The same model is simultaneously a powerful instrument for reading the cultural field and a new actor rewriting it. We return to this tension, formalized as a kind of uncertainty principle, in the closing sections, along with its ethical implications.

## Culture and Cognition in the Age of AI

The interpretive turn in mid-twentieth-century anthropology, led by Geertz, Douglas, Turner, and Schneider, recast culture as an evolving network of publicly shared meanings distributed across actors in time and space. Geertz's *The Interpretation of Cultures* (1973) coined the analytic strategy of "thick description" to unpack the layered and situated meanings of cultural events. Turner's analyses of Ndembu ritual (2016) and Schneider's reading of kinship as a system of symbolic relations extended the program, while Douglas argued in Rules and Meanings that "human thought serves human interests, and therefore carries in itself at any given moment the social configurations of that time and place" (1973: 11). The unifying premise was that culture is a publicly shared, dynamically evolving system of material and symbolic artifacts held together by the interpretive associations linking them. Culture and cognition were theorized as intrinsically linked.

The interpretive tradition supplies the framing within which current AI technologies acquire their cultural-scientific significance (Clifford & Marcus, 2010). We treat culture, with the interpretivists, as a system of meaning at the interface of cognition and sociality. From this premise, three lines of work converge. Microcultures have been historically documented by anthropologists through immersive thick description within small groups and communities. Psychologists, sociologists, and organizational scholars have characterized macro-cultures within and across societies through “thinner” quantitative evidence: surveys, behavioral measures, the anthropological record of artifacts and ethnographies (Roscigno and Hodson 2004; Roscigno, Lopez, and Hodson 2009; Jackson et al. 2019). Cultural evolutionists, beginning from work on the intergenerational transmission of material technologies and language, have built mathematical accounts of how distributions of cultural variants change over time as a function of evolutionary constraints and selection mechanisms (Boyd and Richerson 1988; Henrich and Boyd 2002; Henrich 2016; Mesoudi 2011; Kirby, Cornish, and Smith 2008; Morin 2015).

Across these traditions remains an unresolved question of integration. Macro accounts of cultural form sometimes locate explanatory power in collective behavior alone, treating the personal characteristics of individuals (e.g., their attitudes and beliefs) as socially constructed byproducts (DellaPosta, Shi, and Macy 2015; Macy et al. 2019). Under popular constructivist accounts, processes of social construction are largely unconstrained by micro-cognitive factors, and trajectories of cultural evolution are fundamentally historically contingent and unpredictable. Yet the absence of a measurement framework for the micro–macro interface has meant that this commitment has functioned largely as a theoretical claim rather than robust empirical evidence. Sociological micro–macro frameworks, including Coleman’s (1986) “boat,” Giddens’s (1986) structuration, and Bourdieu’s (1977) habitus, articulate a powerful recursion through scales (see Figure 1). This can stabilize cultural systems, but it can also unleash higher-level processes of rejection or revolution that limit future prediction (Bateson 1935; Graeber and Wengrow 2021). Until now, however, these multi-level models have largely eluded measurement and systematic quantitative testing.

We argue that the conditions for an integrated empirical science of culture and cognition are now in place. AI methods have brought methodological alignment bridging cognitive science and computational social science (Hannan et al. 2019; Mohr et al. 2020), permitting the simultaneous mathematical specification of individual subjectivities, intermediate-scale interactional dynamics, and macro-scale cultural fields. The methods at the heart of this alignment, neural-network representations spaces or embeddings, are continuous with a long cognitive-science tradition that has represented concepts as points or regions in structured geometric spaces (Gärdenfors 2000, 2014; Goldstone and Rogosky 2002; Lupyan and Lewis 2019; Piantadosi et al. 2024), now operating at corpus scales that earlier formal frameworks could not empirically approach.

In what follows we hold three further commitments. First, cultures are high-dimensional multimodal dynamical systems that exist when shared within or across social scales, from individual embodied cognition to societal ideological coordination. Second, the relevant question is not whether to integrate cognitive and social science but how, given that the historical objections to integration have been driven less by principled disagreement than by the absence of a tractable common formalism (Vaisey 2021; Beller et al. 2012; Lizardo 2014; Keen 2014). The next two sections take up these two halves of the integration in turn, the cognitive constraints that culture inherits and the cultural constraints cognition inherits, before turning to the geometric computational framework that allows both to be modeled together.

A third commitment shadows the first two and recurs throughout: the measurement of culture is inseparable from action within it. In Werner Heisenberg's Uncertainty Principle, a signpost of the quantum age, one cannot simultaneously determine both the position and the momentum of a subatomic particle. The more certain one is about where it is, the less certain one can be about where it is going. AI has brought about a structurally analogous trade-off in the study of the social and cultural world. As our instruments for identifying, imaging, and computationally characterizing subjectivities and cultures have grown more precise, counterintuitively, the ability for these instruments (e.g., deep learning embeddings) to forecast the evolutionary trajectory of cultures has worsened; we argue this is because, increasingly, the machinery we rely on to quantitatively describe culture is the machinery now deployed to build agents that act within it and thereby shape it at a large and unprecedented scale. We articulate and explore this Uncertainty Principle for generative AI, and its consequences for cultural predictability and evolution, after establishing the measurement framework it presupposes.

## The Cognitive Foundations of Culture

Cognition contributes to culture a sense of functional integration, a set of interlocking component processes, and a set of productive constraints. Because cognition is typically understood as mental, computational processing in the service of purposeful action, including prediction, decision-making, and problem-solving, it carries with it an implicit orientation toward the goal-directed individual. Insofar as cognition is recognized to contain componential and modularized functions (e.g., representations, learnability, memory), and categorical structure (e.g., clustering and hierarchy), which have been refined through biological and cultural evolution to support prediction and action (Heyes 2018; Tomasello 2009, 2016), so culture inherits these constraints. Progress in measuring and modeling individual cognitive constraints on cultural transmission renders larger-scale cultural trends increasingly predictable.

One metaphor for these constraints is provided by the frame of predictive processing (Friston 2010; Clark 2013, 2016). In this theory, the brain's primary and constitutive aim is to minimize error between its generative model of the world and incoming sensory evidence. It further proposes that cultural artifacts and practices are collectively maintained as scaffolds for prediction and cognitive alignment. They function as shared models that reduce the cognitive cost of inferring others' actions, intentions, and the ambient state of the social world (Veissière et al. 2020). Whether or not this framework is ultimately falsifiable (Bruineberg et al. 2022), we suggest that cultural scaffolds, from institutions to technologies, supervene on cognitive predictions of any type and enable correlated interpretations, via measurement devices or interpretive technologies like LLMs, which further constrain and enable processes of sociocognitive evolution. Within the geometric framework we develop here, a cultural field can be viewed productively as a structured and shared prior over expected meanings, behaviors, and associations, which individual minds inherit and update as they move through it.

Importantly, the constraints imposed by cognition on cultural evolution yield key affordances for macro-level prediction. For example, cultural evolution operates under cognitive constraints that shape the transmission and adoption of ideas. Frequency-dependent selection and fitness landscapes influence how concepts spread through populations, with factors such as learnability, salience, and appeal determining rates of survival. Muthukrishna, Morgan, and Henrich (2016) provide experimental evidence for frequency-dependent mechanisms; in their transmission-chain experiments, minority variants outcompete majority alternatives when they possess greater cognitive fit, constructing selection pressures distinct from conformity bias alone. Guilbeault, Baronchelli, and Centola (2021) further show that even weak cognitive biases toward particular judgments can form a surprisingly strong basin of attraction, predicting the category systems that large human groups form when inventing shared languages for novel phenomena despite considerable cognitive diversity at the individual level. Guilbeault & Centola (2025) replicate these dynamics among content moderation teams evaluating controversial social media images. These results are part of a growing program examining how cognitive constraints generate cultural attractors, which result in predictable convergence dynamics staging the emergence and evolution of socially constructed artifacts and ideas (Sperber 1996; Sperber and Hirschfeld 2004; Claidière, Scott-Phillips, and Sperber 2014; Falandays and Smaldino 2025; Chrisomalis and Miton 2026; Acerbi and de Courson 2025).

A particularly impactful family of cognitive constraints in cultural evolution stems from the embodied character of human cognition. That humans come to know the world through rich experiential, perceptual, and embodied interaction with each other and the environment shapes everything from the structure of concepts to their communication and diffusion through interaction (Barsalou 1999; Lakoff 2012; Shapiro 2019; Pulvermüller 2018; Borghi et al. 2017). This applies not only to concrete concepts but also to abstract ones, commonly conceptualized through metaphors that import concrete, sensible structure into otherwise imperceivable territory, as about other minds or predictions of the future. Recent work confirms that concrete concepts are more stable in their patterns of cultural evolution and more likely to emerge with similar, translatable meanings across contexts and languages (Lewis et al. 2023). Related work finds that measuring the similarities in concepts between cultures is significantly enriched by examining the images people produce when imagining concepts, above and beyond text-based measures of semantic similarity (Pera et al. 2026). The greatest level of cross-cultural image consistency emerges for the concrete and haptically engaging concepts (Lewis et al 2023; Pera et al. 2026). This has direct implications for cultural evolution: shifts in the concrete, embodied means of representation and interaction can produce large-scale influence on the formation of cultural meanings. Guilbeault, Delecourt, and Desikan (2025) show that recent shifts toward visual communication in image- and video-based social media amplify the salience and spread of stereotypes, since images supply both more concrete and perceivable data of associations between demographics and social categories, and more potent and memorable psychological primes. This raises a productive question for research on cultural evolution under the AI turn: by changing the means through which content of various modalities can be created and communicated *en masse*, how will AI technologies reshape the embodied dynamics of meaning-making, and the attractors in concept diffusion that follow?

The multimodal nature of embodied cognition and its interaction with evolving communication technologies is one of the most important frontiers for research at the intersection of AI and culture. While AI systems are increasingly multimodal, evidence suggests that they approach multimodality in ways qualitatively distinct from human cognition (Kadambi et al. 2026; Barrett and Miller 2026). Whether the statistical paradigm driving current AI is capable of learning, and more precisely *experiencing*, perceptions and feelings comparable to those that pervade human embodied cognition is the subject of an active and consequential debate. Bender and Koller (2020) sharpened the issue with a thought experiment about representing the world of the octopus, arguing that a system trained only on text-form linguistic communication, no matter how richly distributional, cannot in principle ground its outputs in the world they describe. Harnad (1990) had earlier framed this as the symbol grounding problem and Glenberg and Robertson (2000) provided experimental evidence that purely distributional accounts fail to capture what humans understand about action-related sentences. Counterarguments hold that distributional learning over multimodal corpora may approximate aspects of grounding (Pavlick 2023), and that the relevant question is empirical rather than principled (see Kadambi et al. 2026 for recent review).

The empirical evidence so far suggests that current AI systems remain meaningfully different from human cognition in their handling of embodied content, though the divergences are partial and selective. In the illustratively simple case of color, uniquely salient in primate vision and pervasive in human abstract cognition (Guilbeault et al. 2020; Desikan et al. 2020), vision models consistently fail to learn color in service of classification, and when they do retain it, their similarity judgments and semantic associations deviate systematically from human color perception and synesthetic reasoning (Nadler, Darragh-Ford, Srinivasa Desikan et al. 2023; Bujack et al. 2022; Nadler, Guilbeault, Ringold et al. 2025). This suggests deeper differences in how embodied associations factor into abstract interpretation. These cases sit within a broader literature which finds that neural network models reproduce human behavioral signatures in some domains, while diverging sharply in others (Lake, Ullman, Tenenbaum, and Gershman 2017; Bowers et al. 2022; Binz and Schulz 2023; Frank 2023; McCoy, Yao et al. 2024, Kargupta et al. 2025).

We do not take the embodied gap to settle the question of whether the statistical paradigm is irredeemably insufficient, and we avoid confident claims in either direction. What we do hold is that the gap presents falsifiable predictions. If distributional training over increasingly multimodal corpora is sufficient for cultural modeling, then domains with rich non-linguistic structure should yield to improvements with appropriate training data. If not, those same domains should remain stubbornly resistant. Recent work has taken the question on directly. Augmenting deep learning models with explicit knowledge of sensory experiences—for instance, the chemical composition of molecules (Makri et al. 2026)—improves the ability of large language models to generate coherent predictions about smell, despite language data on olfactory-related metaphors being comparatively underdeveloped. Neurosymbolic approaches that incorporate explicit knowledge structures relating to human categorization and perceptual experience (Barrett and Miller 2026; Guilbeault, Caplan, and Yang 2026; Roll et al. 2026) may yield further progress. Whether embodied mechanisms of cognition, such as complex analogical reasoning, synaesthetic associations, affective cognition, and conceptual leaps that elude current LLM architectures (Duede and Evans 2021; Brucks and Levav 2022; Zahavy 2026) turn out to be necessary for the predictive performance we describe in subsequent sections is, we believe, one of the central empirical questions of the next decade.

Even so, the embodied gap does not currently prevent embedding methods from providing significant utility as a measurement technology in modeling the multimodal structure of concepts. Consider the multimodal, multilayered structure of cultural constructs relating to sports: a particular team is associated not only with semantic knowledge in textual form but also with visual symbology of logo and color, auditory theme songs, and a gambit of immersive ritualized behaviors and gustatory experiences that constitute the full contextual representation of the team in the lives and minds of its fans. As multimodal models progress in their ability to embed and associate cultural objects across modalities, they unlock new capabilities for characterizing the multimodal nature of human concepts and behaviors in embodied cultural contexts.

This progress is unlikely to follow a unidirectional, bottom-up logic in which cultural constructs are reduced to the sum of their multimodal associations. A central premise of this article is that AI is bringing about a computational paradigm for cultural analysis that reveals and embraces its intrinsically multiscale nature: macro constructs such as ideologies, markets, and nations are themselves represented and modeled as defining the conditions under which socially embedded behaviors, associations, and interpretations unfold. To complete the framework, we now return to the cultural foundations of cognition.

## The Cultural Foundations of Cognition

Sociological accounts of cognition have long insisted that the categories and concepts through which individuals perceive, classify, and reason are neither private inventions nor neutral reflections of the world, but are instead collectively produced and institutionally sustained. Durkheim early made the argument in *The Elementary Forms of Religious Life* (1912) that basic categories of understanding, including self, time, space, causality, and classification, originate in the organization of social life. He distilled that intuition within questions like which came first, the set-theoretic notion of a group as a collection of elements, or the concrete notion of a social group with members from which the abstract mathematical form was analogically derived?

Mary Douglas sharpened this view in *How Institutions Think*, arguing that human conceptual systems and organizational structures are defined by and justified through fundamental analogical relationships, grounding abstract organisational structure in concrete logics of the natural world, such as grounding theological hierarchy in a view of a natural hierarchy among animals. Taking this further, Bourdieu's notion of *habitus* describes how durable dispositions of perception and judgment may be inculcated through socialization into class and field positions, such that reasoning styles carry the fingerprint of one's location in social structure (Bourdieu 1977, 1986, 1987; Lizardo 2006; Goldberg 2011). The cognitive sociology developed by Zerubavel, DiMaggio, Cerulo, and others has similarly mapped how classifications, attention, memory, and analogical reasoning are socially patterned within and across "thought communities" (Zerubavel 1997; DiMaggio 1987, 1997; Cerulo 2001). Swidler expands the role of human agency within cultural fields by recasting culture as a differentiated toolkit of strategies, frames, and justifications from which actors may selectively and strategically draw (Swidler 1986). More recent dual-process and practice-centered accounts reconcile these approaches by distinguishing between fast, automatic, schema-driven cognition that carries most everyday interpretation and slower, discursively elaborated reasoning people mobilize when asked to justify themselves (Boutyline and Soter 2021; Lizardo 2014), yielding predictions about which forms of cultural influence are accessible to introspection. Other forms of cultural influence must be recovered indirectly, including via ambient, multimodal cultural influences that embedding-based methods are poised to capture.

Anthropological perspectives complement these sociological claims by documenting the variability of conceptual schemes, interpretive frames, and inferential habits that human communities develop. Furthermore, they theorize the structural mechanisms through which such schemes become cognitively binding. Building on Geertz's interpretive program, cognitive anthropologists have shown how shared cultural models mold everyday reasoning about kinship, morality, illness, and the self (Shore 1996; Lévi-Strauss 1963; D'Andrade 1995). Parallel traditions have developed similar intuitions about how cognition is distributed, reinforced, and extended through tools and the design of cultural environments (Miton and Jackson 2026; Hutchins 2014; Heersmink 2017). Hutchins's analyses of navigation and aviation demonstrate that complex cognitive tasks are routinely distributed across people, artifacts, and institutional routines, such that the relevant unit of analysis for many forms of thinking is a sociotechnical system rather than a singular mind (Hutchins 1996). Clark and Chalmers (1998) argued in turn that this distribution is not merely operational but constitutive: the boundary of the cognitive system extends into the artifactual environment whenever those artifacts play the right functional role. Psychological research has increasingly converged on complementary insights through the influence of Vygotsky (1978) and Luria's (1976) cultural-historical psychology, and the explosion of work documenting ways sociocultural contexts impact individual and group psychologies (Markus and Kitayama 1991; Nisbett, Peng, Choi, and Norenzayan 2001; Masuda and Nisbett 2001; Nisbett 2003; 1994; Henrich, Heine, and Norenzayan 2010; Heine 2020; Boroditsky 2001; Lupyan and Dale 2010; Majid et al. 2018).

Across these literatures, a cumulative picture emerges in which cognition is conditioned by the cultural environments in which minds develop and operate, a picture we argue is newly tractable for computational investigation once culture is represented as a learnable, high-dimensional structure whose geometry can be related directly to the subjectivities embedded within it. Cultural evolution research further argues that the distinctive cognitive profile of our species rests on capacities for shared intentionality, high-fidelity social learning, and cumulative cultural ratcheting that allow successive generations to inherit and refine conceptual tools, norms, and technologies they could not individually invent (Boyd and Richerson 1988; Tomasello, Carpenter, Call, Behne, and Moll 2005; Tomasello 2009, 2016; Henrich and Boyd 2002; Henrich 2016; Csibra and Gergely 2009). The 'cognitive gadgets' account extends this lineage, emphasizing the causal role of tool design in neurocognitive development. For example, reading and writing recruit a specific, conserved brain network and constitute a synesthetic leap of embodied cognition, rendering the sounds of oral language communicable as visual symbols (Heyes 2018).

Taken together, these strands establish that much of what is cognitively *inside* individuals is the sedimented residue of historically specific collective practices and technologies. This motivates the effort to represent cultures themselves, and not merely individuals, as structured high-dimensional systems amenable to computational analysis. More broadly, insofar as embedding-based models of science are increasingly able to predict scientific and technological inventions, they may provide a vehicle for anticipating future constraints on cultural evolution stemming both from how cognition constrains culture and how cultural artifacts shape cognition (Rahwan, Shariff and Bonnefon 2025). As Alan Kay put it, the best way to predict the future is to invent it.

Ongoing work in computational social science is identifying new ways through which culture both transcends and impacts individual cognition through emergent interactions between representations and people. González-Avella, Cosenza, and San Miguel (2014) demonstrated how collective culture emerges from the network interactions between individual cognitions and the environment, generating complexity irreducible to individual cognition. A second-order role of culture involves awareness of diversity, conflict, and "counter-cultural" understandings, visible only against the backdrop of shared institutions. This recognition of multiple perspectives enables new identities to scale from individuals to collectives, allowing collective cognition and action to adapt to new circumstances over time. Shi and Evans (2023) documented these evolutionary processes in their network analysis of semantic embeddings across scientific disciplines: interdisciplinary knowledge transfer occurs through reconfiguring conceptual relationships rather than simple adoption. This suggests that cultural evolution may operate through the transformation of meaning systems via social interaction rather than mere replication, especially in cultural domains attuned to innovation like art and science.

In sum, advancements across the social and cognitive sciences provide increasingly coherent and cumulative evidence that individual cognition and collective culture mutually constrain and influence each other. The relationship is not deterministic, but predictable evolutionary dynamics and mechanistic constraints make a science of the cognitive–cultural interface possible. A key challenge in advancing this science will be developing mathematically and computationally tractable models of culture as a high-dimensional system spanning both micro and macro scales. The remainder of the article assembles developments for this purpose, beginning with a concise statement of the framework that the foregoing discussion converges upon.

## Culture as a High-Dimensional Multimodal Field

Weaving together threads from the previous two sections, we treat cultural content, ranging from texts and images to practices, routines, and behavioral trajectories as positions within a high-dimensional, multimodal semantic space. In this view, an individual identity is a subjective *trajectory* through that space: a locally dense cloud of association, attention, and expectation that each person carries as a function of their experiences, biography, training, language and other patterns of cultural exposure (Goodenough 1981, D'Andrade 1995). In panel A of Figure 1, we illustrate how patterns of shared cognition vary across levels of analysis. Here we illustrate culture across four levels. Level 1 represents individual cognition and subjectivity that fuses into stable but distinctive identities and worldviews sometimes described as idio-cultures[3] captured by points within the geometric space. We sidestep the question of whether an idio-culture is a culture at all—a category error if cultures are shared systems of public symbols (Wittgenstein 1953; Geertz 1973; Lewis 1969; Kripke 1982), or the degenerate limiting case of a distribution of representations on Sperber's (1996) epidemiological account. In this, we follow cognitive anthropologists who posited a level below culture (e.g., Goodenough's (1981) "propriospect") assembled from individual exposure to many cultural traditions (D'Andrade 1995), which reflects our trajectories through Levels 2–4.

Level 2 involves shared micro-cultures that emerge through interactions within social networks and across families, friendships, and local communities. Level 3 constitutes meso-scale cultural fields that cover entire organizations (e.g., Apple, Google), markets (e.g., the art market), and social movements (e.g., Black Lives Matter). Level 4 captures the macro-cultural zeitgeist and ideology immanent within larger times and places, like Weimar Republic Berlin of the 1920s or the United States of the 1960s.

3 We note that Gary Alan Fine uses "idiocultures" differently as the culture of a small interacting group (1979), but this usage lies at variance with the prior use of *idiolect* as a private language (Wittgenstein 1953), *idiodynamics* as personality dynamics (Rosenzweig 1951), and *idioverse* as an individual worldview (Rosensweig 1958).

The diagram in panel A highlights two distinctive properties. First, each line traces an individual identity and the subjective perspective that comes from its distinct position within cultural fields. This renders the question of what level cultures are most "real" an empirical (and geometric) one: at the level in which they are most distinctively clustered, where clustering is a function of closeness, coherence, or density within cultural fields, and separation or distance between them. These properties determine the degree of culturally shared and unshared patterns of cognition. Second, the figure highlights how cultures at lower and higher levels condition and effectively compete with one another. Sharing the same language or ideology constrains the cultural distance possible at lower levels, just as the distinctiveness of lower-level micro-cultures limits the sharedness possible within a higher-level macroculture. The coherence and separation of cultures at one level makes it harder for the manifestation of coherence and separation at another level.

Panel B highlights two additional properties of culture. First, cultures emerge, evolve, and recede over time. At one point, $t = 5$ within the diagram, a large-scale macro culture is coherent when at other times, $t = 10$, it fragments into meso- or micro-cultures. Consider a nation whose culture coheres in wartime around shared symbols and significations, then weakens and intermingles with neighboring cultures in peace, yielding a smoother and more differentiated field. Second, a culture may be understood as an "attractor basin" whose contours shape the local search behavior of individual human trajectories that pass nearby. People whose identities intersect within a cultural field come to adopt or reject subjective perspectives shared within that field.

The goal of this framework is not to reduce culture to these computational primitives alone, but rather to formalize a wide range of complex cultural dimensions through an integrative computational lens capable of characterizing measurable, interactive and co-constitutive relations between micro and macro scales. As we discuss below, our view does not advocate for a naive determinism under which, given enough data and computing power, computational analysts will be able to accurately model the full trajectories of subjectivities and entire cultures. Instead, as we emphasize throughout, the co-constitutive dynamics between cognition and culture introduce feedback loops unlikely to reduce to simple linear dynamics, including Bourdieu's habitus, Giddens' structuration, and Bateson's schismogenesis articulated earlier. Nevertheless, we maintain these co-constitutive dynamics are increasingly measurable via recent computational techniques like geometric embeddings, yielding improvements to prediction on the one hand, and the discovery of new and more important or explanatory feedback loops driving cultural evolution on the other.

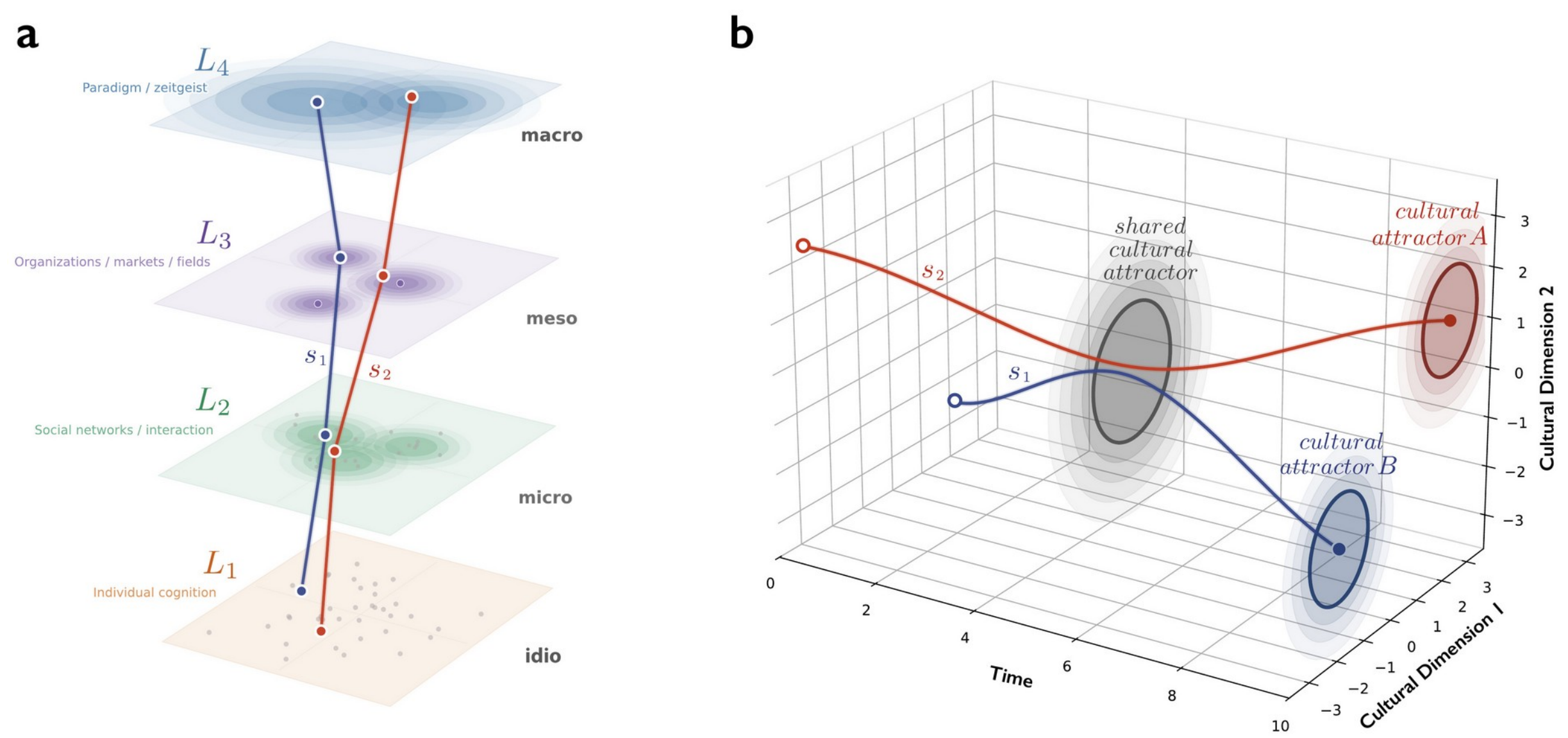

**Figure 1. Identities as subjective trajectories through measurable cultural fields. (a) Across scales**. Two identities (red, blue) thread four nested embedding spaces: idio-cultures of individual cognition and worldview (L1), micro-cultures emergent in interaction and social networks (L2), meso-cultures of organizations, markets, and movements (L3), and macro-cultural zeitgeists, ideologies, or paradigms (L4). Shaded densities are cognitive attractors built from aggregated trajectories; the same vector geometry applies at every level. **(b) Through a drifting field.** Both identities (open circles, start; filled, end) are first drawn toward a shared attractor (e.g., a wartime zeitgeist), through which independent minds converge—making simultaneous discovery predictable—before separating into distinct attractors (A, B) as the attractor landscape itself shifts.

Three commitments follow. First, our framework is *scale-invariant in its mathematics*: the same vector geometry that effectively measures individual concept structure (Piantadosi et al. 2024; Gärdenfors 2014) also captures organizational discourse (Srivastava, Goldberg, Manian, and Potts 2017; Lix, Goldberg, Srivastava, and Valentine 2022) and macroscopic ideological structure (Kozlowski, Taddy, and Evans 2019; Waller and Anderson 2021). What changes across scales is the unit of analysis, not the geometric mode of characterizing identities and relations. Second, the framework is *modality-extensible*: any source of culturally meaningful data—text, image, audio, video, behavioral logs, biological measurements—can in principle be embedded into a shared representational space (Adukia et al. 2023; Guilbeault et al. 2024; Savcisens et al. 2023); see Figu 2. Third, the framework is *generative in two senses*: it permits the simulation of behavior from particular cultural positions (Argyle, Busby, et al. 2023; Park et al. 2024; Hewitt, Ashokkumar, Ghezae, and Willer 2024), and it permits the workable prediction of how the field itself will evolve under specifiable conditions (Belikov et al. 2022; Shi and Evans 2023; Kim, Kojaku, and Ahn 2026). While the large-scale cultural conditions amenable to useful prediction may be limited in scope, the growing demonstration of predictability across a variety of contexts marks a qualitative shift in the capacity for AI to both model and causally impact trajectories through cultures and the evolutionary trajectory of cultures. This unlocks a new micro–macro feedback loop with potentially unprecedented evolutionary dynamics. Note that the two generative senses are not independent capacities but one machinery read in opposing directions: the mirror that simulates behavior from a cultural position is the engine that can act from that position. This identity is what gives the framework its uncertainty structure, developed below. Every gain in the precision with which a cultural position is fixed equips agents, human and artificial, to act on that position, altering the very trajectory the measurement would otherwise have licensed us to predict.

We turn next to the technical apparatus that makes these commitments operational, then to substantive findings that motivate them.

## Modeling the Multiscale and Multimodal Nature of Culture

The technology of contextual embeddings, the representations within modern deep neural networks, have opened new possibilities for cultural analysis. They provide a means to measure subjectivities in terms of how the meanings of words change with surrounding social context (including who is speaking), and to compare the semantic universes of individuals and contexts through geometric and graph-theoretic measures of similarity in how associations are constructed. These achievements are not entirely dependent on text and can in principle include any format of cultural information, where cultural objects can be quantified by their adjacency to other such objects within cultural contexts and feature spaces. Artificial neural networks (ANNs) extend the framework beyond textual data by providing a generalized framework for including diverse modalities of culturally rich data—audio, images, video, behavioral recordings—into an integrated, multimodal embedding space within which the respective contributions of these modalities to subjectivities can be examined.

We illustrate this in Figure 2A, where the modern concept of a nation is refracted through text (e.g., constitutions, government websites), images (e.g., flags, monuments), audio (e.g., anthems, speeches), video (e.g., propaganda, armed services recruiting clips), networks (e.g., diplomatic ties, trade), and behaviour reflecting shared values. The result is a cultural field of distinct nations, each represented by a Greek letter, which vary in their density, separation, and overlap.

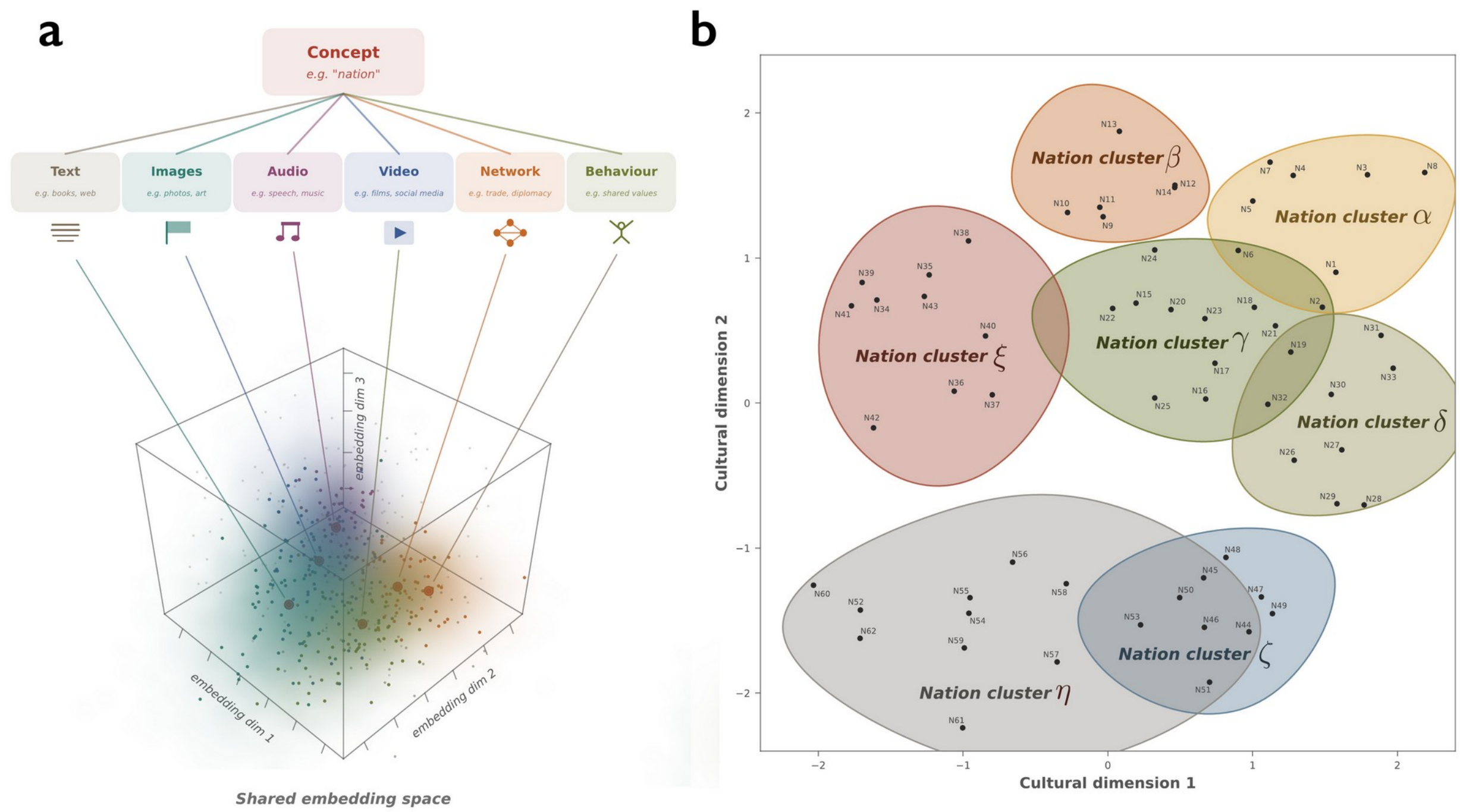

**Figure 2. Multimodal cultural data mapped into a single shared geometry. (a) Aggregation.** A single concept (here, 'nation') expressed across six modalities—text, images, audio, video, networks, and behaviour—is projected into one vector space (colored discs, modality regions; red points, centers), making cultural content beyond text directly comparable and combinable. **(b) Clustering.** Seven nations (signified *α*–*η*; shaded ellipses) positioned in a two-dimensional time slice fall into latent clusters of cultural artifacts (N1-62), some well separated, others overlapping: the geometry that captures individual concepts also resolves large-scale ideological structure.

In its most recent form, this technology has produced generative algorithms, like the pre-trained transformer-based ChatGPT, that harness contextual embeddings to produce novel combinations of text in ways that are useful insofar as they are sensitive to cultural information. Context-sensitive embeddings differ from earlier static embedding approaches by dynamically reweighting the embeddings associated with particular words or larger bodies of text as a function of surrounding context. As a result, the embedding for a given word like "American" can change as a function of whether it appears in left- or right-leaning text. This innovation has led to the digital twin framework, whereby generative transformers are prompted to produce text, judgments, or behavioral inputs as if operating from the subjective viewpoint of a particular kind of person in a particular cultural context, such as a Republican attempting to change the opinion of a Democrat on a contentious issue (Argyle, Busby, et al. 2023; Kozlowski, Kwon, and Evans 2024). ANNs together provide a suite of new tools for capturing both sides of the micro–macro interface and their interaction by approximating micro-subjectivities in terms of how particular kinds of people, or even specific individuals (Park et al. 2024; Kim and Lee 2026) interpret and produce meanings in unique, contextually sensitive ways, while also measuring how the broader landscape of meaning shapes interpretations and the communicative affordances available to actors at a given time and place. The remainder of this section provides an overview of recent uses of AI to measure and illuminate this interface.

### *AI Models of Subjectivities in Context*

Given the enormity of behavioral and language data online, ANNs make it possible not only to form embeddings that capture the landscape of meaning for the entire internet but also to build localized embeddings that reflect the subjectivities of particular types of agents within this landscape. Researchers have developed several ways of adapting language models to capture the specific styles of association, speech production, and semantic interpretation that characterize particular agentic viewpoints.

In generative transformer models, representing subjectivities can be approximated through prompting. Researchers can prompt an LLM to produce or evaluate text from the perspective of a particular kind of person. Recent work shows that when LLMs are asked to complete canonical survey tasks and game-theory experiments from the perspective of particular agents (defined by gender, race, income, partisanship, etc.), they replicate the corresponding participants' responses in a way that captures over 80% of the variance, providing a robust framework for predicting behavior (Hewitt et al. 2024; Park et al. 2024; Argyle, Busby, et al. 2023). Recent approaches to mechanistic interpretability of LLMs to facilitate direct, geometric steering of model internals to modulate perspectives within the embedding space by discovering the embedding layer(s) within the model most linearly associated with a given cultural or psychological perspective (Kim, Evans, and Schein 2025; Kozlowski and Evans 2025; Chen et al. 2025).

Alternatively, LLMs can be fine-tuned on data from particular individuals or groups, reweighting embedding-space distances toward the new co-occurrences, either through full retraining (e.g., bioBERT, legalBERT; Lee et al. 2020; Chalkidis et al. 2020), parameter-efficient adapters that bias weights toward a distinctive cultural voice (Hu et al. 2021; Dettmers et al. 2023), or reinforcement learning toward a given cultural constraint (e.g., seeming "helpful") or subjective perspective (Christiano et al. 2017). Recent work also suggests enduring limitations in the ability for LLMs to simulate individual variation of interpretation to arbitrary levels of precision, however; the dependence on LLM representation on aggregate statistical representations of people and culture lead to statistical attractors, defined at a global level, that struggle to meaningfully reproduce local subjectivities (Wang et al. 2026; Zhao, Guilbeault & Goldberg, 2026).

There are numerous ways language models are being used to theorize and measure micro-psychological factors mediating cultural dynamics. This capacity rests on the result that distances among concepts in embedding space perform well at predicting people's judgments of semantic relatedness in psychological experiments (Grand, Blank, Pereira, and Fedorenko 2022). One promising line of work concerns the identification of psychological biases that mediate the kinds of semantic associations and judgments people form, with measurable behavioral consequences. By comparing the structure of semantic associations in ANNs trained on different languages, researchers have identified how semantic associations for concrete concepts are more stable and easier to translate across languages than those for abstract concepts (Lewis et al 2023), consistent with individual-level psychological evidence of greater subjective variation in the interpretation of abstract versus concrete concepts (Wang and Bi 2021). Similar methods have identified patterns of cultural variation in the semantic associations for emotions (Jackson et al. 2019). A complementary finding is that independent LLMs themselves appear to exhibit convergent patterns of evolution toward similar shared embedding spaces, with the probability of independently converging to shared representations of concepts (Huh et al. 2024) and values (Mazeika et al. 2025) rising with model size, which is consistent with the known effects of population scale on the emergence of shared meanings in human groups (Guilbeault et al. 2021).

The ability for LLMs to capture how individuals interpret everyday concepts also extends to capturing how individuals make sense of culturally contextual phenomena. Recent work shows that LLMs are effective at capturing the semantic structure of stereotypes and at predicting stereotypical interpretation patterns, due in part to their reliance on learning aggregate statistical representations from human text, where such statistical aggregation processes have long been viewed as the underlying psychological driver of stereotype formation among humans (Allport 1954; Bai, Fiske & Griffiths 2022; Gulbeault, Delecourt & Srinivasa Desikan 2025). Kozlowski, Taddy, and Evans (2019) developed the *Geometry of Culture* method, which generalizes to LLMs (Kim et al 2025), and extracts cultural dimensions from embedding spaces and positions concepts within them to identify their stereotypical associations. They demonstrated that semantic associations surrounding common domains like sports and music are infused with class-related stereotypes (golf is more associated with affluence, boxing with its absence) and that while the markers of class shifted amid the economic transformations of the twentieth century, the basic cultural dimensions of class remained remarkably stable.

A number of other studies, including Lewis and Lupyan (2020) and Bailey, Williams, and Cimpian (2022), have developed similar approaches to recovering gender stereotypes at scale. Bailey and colleagues find that the abstract concept of "person" is more associated with men than women in large-scale patterns of language use on the internet. Others use related approaches to compare gender embeddings across historical datasets to identify ways in which stereotypes are fading and, in some cases, intensifying (Charlesworth, Caliskan, and Banaji 2022; Jones, Amin, Kim, and Skiena 2020; Caliskan, Bryson, and Narayanan 2017; Caliskan et al. 2022; Garg, Schiebinger, Jurafsky, and Zou 2018). Some analyses go beyond documenting stereotypes by using embeddings to reveal novel aspects of stereotypical associations, as in Card et al.'s (2022) demonstration that biases toward immigrants often take the form of implicit, dehumanizing metaphorical associations with animals or objects, such as "cargo".

These insights are not limited to textual data. Waller and Anderson (2021) used the Geometry of Culture method with patterns of co-participation on Reddit to discover dimensions of cultural association underlying online polarization. Guilbeault et al. (2024) developed a method for integrating large-scale image data into the same framework, finding not only that gender associations across categories are highly correlated between online images and text, but also that the gender biases presented in images are consistently stronger and more psychologically impactful than those captured by corresponding text models, providing new insight into the irreducibly multimodal cognitive structure of latent cultural stereotypes. Guilbeault, Delecourt, and Srinivasa Desikan (2025) build on this work by adapting this approach to represent intersectional biases, such as gendered ageism, across text, images, and video. They further show how using embedding models to identify stereotypical associations in large-scale multimodal datasets can yield robust predictions about biases in the behavioral responses of both humans and LLMs, for example, by predicting biases in LLM's generation and evaluation of resumes. These findings exemplify the growing capacity for AI-extracted aggregate patterns in macro social data to make testable and accurate predictions about the cognitive structure and behavioral patterns characterizing humans and LLM-agents at the micro level.

### *AI Models of the Macro Social Contexts Shaping Subjectivities*

The ability of LLMs to robustly capture semantic judgments in particular contexts opens the possibility of generalizing beyond modeling individual perspectives into modeling larger-scale cultural trends. Contrary to the Saussurean ideal of arbitrary semantic association, this work reveals robust predictability in the new semantic associations that people form in response to cultural events and innovations. Sociological research has often argued that the effects of partisan identity on people's cultural associations can be arbitrary, path-dependent, and unpredictable (DellaPosta, Shi, and Macy 2015; Macy, Deri, Ruch, and Tong 2019). Yet a recent study found such predictability empirically: Kozlowski, Kwon, and Evans (2024) used a large language model trained on pre-COVID data (GPT-3) to predict the attitudes and arguments that Republicans and Democrats produced toward vaccine mandates in the COVID era. Current efforts are predicting new conceptual innovations in real time, with surprising results, including the finding that prescient ideas often emerge from the cultural periphery in terms of which actors hold influence in society (Vicinanza, Goldberg, and Srivastava 2023). This is consistent with theoretical findings from research on "complex contagions", where cultural forms like ideologies are only "caught" through multiple exposures, that innovations are often most likely to emerge and spread from the periphery of social networks (Guilbeault & Centola 2021; Tschofenig & Guilbeault 2026).

Importantly, these predictions do not require LLMs to precisely capture specific individuals' associative reasoning in particular contexts to predict cultural trends. Research suggests that LLMs are limited in reproducing the true heterogeneity of individuals' opinions and interpretations (Wang, Morgenstern, and Dickerson 2025; Santurkar, Durmus, Ladhak et al. 2023). This is because LLMs are designed to compress aggregate patterns that, by definition, generalize across individuals at a coarser-grained level than is required to retain specific subjectivities (Zhao, Guilbeault & Goldberg, 2026). The heterogeneity problem can be partially addressed through improved prompting and steering: LLMs' default responses often struggle to reproduce localized and heterogeneous cultural associations and idiolects, but recent work shows they are coherently responsive to prompts and vector steering designed to convey and elicit the localized cultural forms of specific groups (Veselovsky, Argin, Stroebl et al. 2026).

While solving the heterogeneity problem would enhance the utility of LLMs as social simulators, it is not necessary for LLMs to serve as robust simulators and predictors of appropriately scoped macro dynamics. One reason is that cultural phenomena themselves characteristically aggregate over and collapse the underlying heterogeneity from individual subjectivities, much of which remains a black box even to human observers within the culture. Despite widespread cognitive diversity, cultural trends often reflect weak cognitive preferences that become amplified as people communicate in large populations, leading to stable and replicable trajectories in the emergence of new associations at the macro scale (Guilbeault et al. 2021; Schelling 2006; Acerbi and de Courson 2025). These biases need not be universal or widespread. Tipping points in cultural norms can be activated by coordinated preferences among a quarter of the population or less (Centola, Becker, Brackbill, and Baronchelli 2018; Iacopini, Petri, Baronchelli, and Barrat 2022), though thresholds remain methodologically debated, and preferences shared by less than 20% of a population can lead large, separate groups to independently form similar meaning systems (Guilbeault et al. 2021). The upshot is that while LLMs may be limited in capturing the full breadth of variation in human subjectivity and association, they effectively estimate large-scale cultural attractors that aggregate over this variation and constrain cultural dynamics.

That LLMs, by design, capture the most stereotypical human features and preferences, while undoubtedly problematic in many social contexts, may allow them to accurately model the weak biases that shape large-scale cultural evolution. LLMs achieve this by capturing the aspects of associative reasoning predicted by human identification with particular social groups, such as biases in their reasoning about COVID vaccines as predicted by partisan identities (Kozlowski et al. 2024), which are known to play a significant constraining role on beliefs. Some identities are expectedly more constraining than others. Contemporary U.S. partisan identities are especially entrenched, while other identities, such as whether someone is a birder or prefers coffee to tea, play a weaker role in shaping semantic associations, making heterogeneity along such social dimensions harder to capture via the automated learning of aggregate cultural trends. A general expectation is that LLMs will be effective at modeling existing and novel semantic associations that are heavily constrained by established categories and frameworks. For example, work shows that LLMs are highly effective at learning how to categorize novels and films into genres, including complex genre fusions (Le Mens, Kovács, Hannan, and Pros 2023a, 2023b), allowing them to develop robust predictions of how humans will categorize genres of literature pieces never seen before.

Some contexts of cultural innovation are more amenable to prediction than others. Some exhibit more stochasticity and recursive dynamics, leading to pronounced patterns of path dependent social construction. This highlights an important ground for theoretical work in the new science of culture: delineating the cognitive and social features that determine how predictable they are according to associated AI models of people and their interactions. Regions of the sociocultural association space that are predictably stable indicate the emergent or institutionally reinforced presence of culture.

Language models also afford insight into the unique associative and communicative patterns that characterize individuals and impact their behavior in specific social contexts, such as organizational settings using email and team slack data (Aceves and Evans 2024; Corritore, Goldberg, and Srivastava 2020; Lu, Chatman, Goldberg, and Srivastava 2023; Yang, Anshuman, Goldberg, and Srivastava 2025). Embedding cultural data within neural networks enable representation of how micro-subjectivities are situated within global, macro-cultural landscapes that mediate and shape how they interact. By embedding the entire history of scientific publications, it is possible to identify and even anticipate which scientists are positioned to achieve interdisciplinary breakthroughs, because such scientists are geometrically embedded at the intersection of distinct, well-defined thought communities (Belikov, Rzhetsky, and Evans 2022; Shi and Evans 2023; Sourati and Evans 2023; Hofstra et al. 2020). Scientific innovation thereby emerges as an intrinsically collective enterprise, arising through surprising connections among communities across a field of theories, arguments, technologies, and institutions. Its single-embedding representation reveals how fields evolve through interaction between individual scientists' subjectivities (Cao, Pan, and Evans 2025) and the broader ecology of scientific practice (Shi and Evans 2023).

By capturing both individual positions within a high-dimensional cultural space and how the global structure of culture unfolds over time according to shifting positions among concepts and artifacts,

AI methods create new possibilities for probing both the past and future of culture that overcome longstanding challenges in historical analysis (Jackson et al. 2022). It is now possible to simulate past human perspectives, ideally by training language models exclusively on past data (Levine, Duvenaud, and Radford 2026) or prompting/steering models to communicate through past perspectives (Vicinanza, Goldberg and Srivastava 2023), allowing for approximate exploration of historical subjectivities. As Varnum, Baumard, Atari, and Gray (2024) argue, "historians have often been limited to relatively indirect proxies, such as private letters, individual portraits, philosophical treatises, or epic poetry, frequently analyzed using qualitative approaches. Historical LLMs could complement such efforts by enabling scholars to simulate and generate somewhat more direct quantitative data on traits, attitudes, and behavioral tendencies" (Langlais 2023; Chen 2024; Chen, Li, Li, and Atari 2024). This may open new doors for identifying patterns of cultural variation unique to specific cultures and historical moments, along with those more broadly shared.

## Mapping the Dynamic Nexus of Culture and Cognition

If we take seriously the proposition that cultural content can be represented as positions within a high-dimensional, multimodal embedding space, then human subjectivities can be characterized as trajectories through that space. Such trajectories comprise locally dense clouds of association, attention, and expectation each person carries as a function of their biography, language, training, and patterns of cultural exposure. On this view, a characteristic feature of subjectivity is how it sees and searches through neighboring cultural worlds, probing the local neighborhood of meanings for useful combinations, analogies, and reinterpretations. This is the cognitive face of what Stuart Kauffman calls exploring the *adjacent possible* (Kauffman 2000): the set of configurations not yet actual but are reachable via a single combinatorial step from what already exists. Local combinatorial steps within complex, high-dimensional semantic spaces can yield cultural "exaptations" or symmetry-breaking transpositions that alter the parameters of the possibility space.[4] At the same time, position within the adjacent possible renders individual search within high-dimensional spaces somewhat predictable. Recent work in the science of science supports this: embedding the search behavior of scientists in high-dimensional spaces allows properties of their location in the geometry of culture to predict which ideas they will combine and which combinations will be impactful (Belikov et al. 2022; Shi and Evans 2023; Sourati and Evans 2023;

4 Exaptation was initially coined in the context of biological evolution as the process in which a biological trait or tool is transposed to take on new functions that differ from the one it originally held (Gould and Vrba 1982).

He-Yueya et al. 2026). By mapping the conceptual vicinity of existing cultural elements, one can identify the reachable combinations that constitute the feasible frontier of innovation. While such conceptual exaptations may permanently alter the future trajectory of culture, exemplifying path dependence, the temporally local processes driving these breakthroughs can be robustly characterized and predicted.

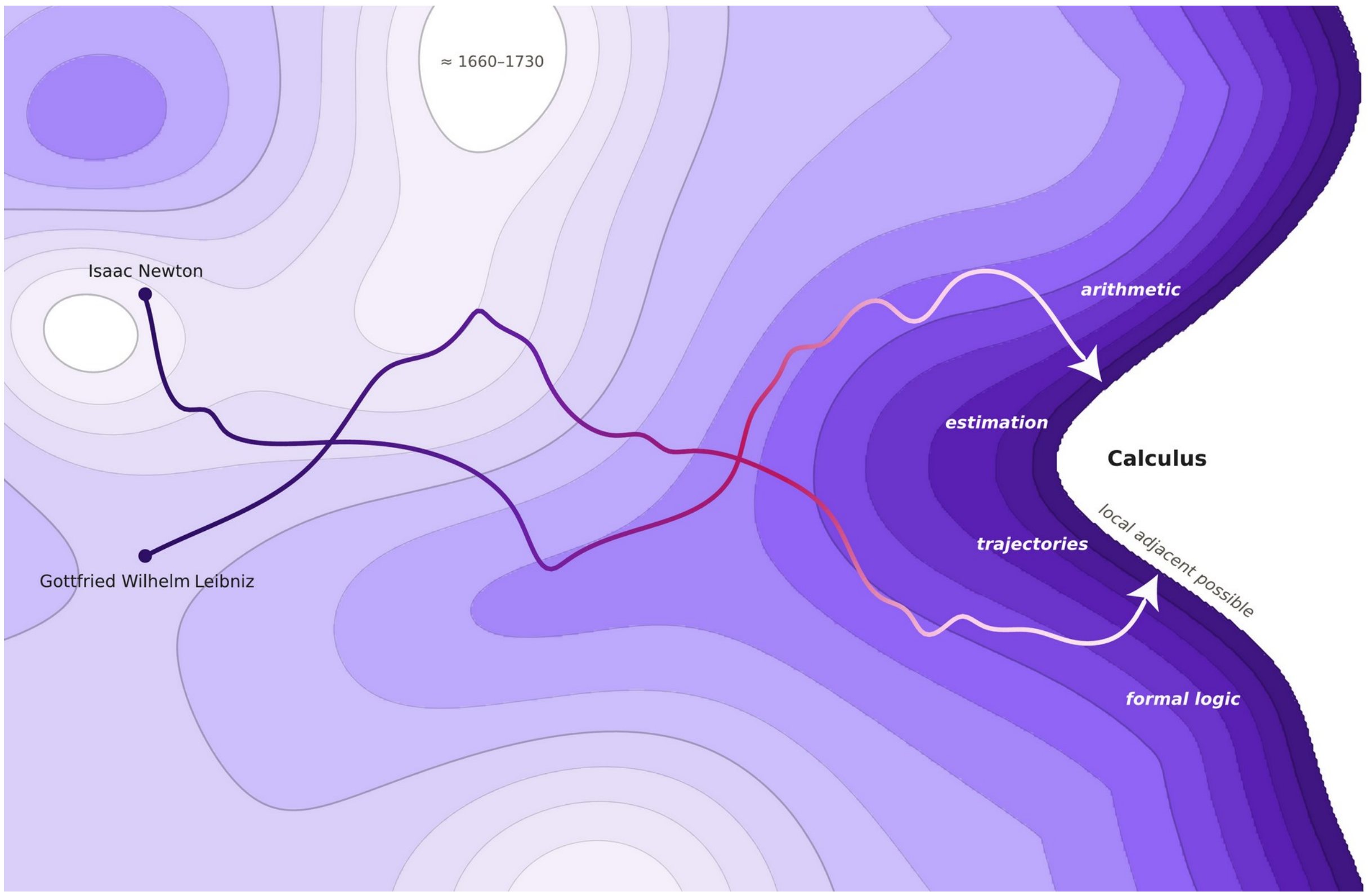

**Figure 3. Simultaneous discovery as convergence through a shared cultural field.** Newton and Leibniz (≈1660–1730) each begin within a small local adjacent possible and trace trajectories (violet) through shared problems of formal logic, arithmetic, and estimation toward a common attractor: the calculus they simultaneously discovered. Because cultural fields constrain where cognitive search unfolds, and are now measurable through embedding geometries, such convergence of independent minds becomes more predictable.

For all the creativity individual creators are capable of, including recombination, exaptation, and analogical leaps of imagination, the historical record contains a stubborn regularity: highly original individuals, working independently and sometimes without knowledge of one another, repeatedly arrive at nearly identical discoveries at nearly the same moment. Newton and Leibniz independently formulated the calculus in the 1670s–1680s; Darwin and Wallace converged on natural selection in 1858; Bell and Gray filed telephone patents on the same day in 1876; the deep-learning architectures that came to dominate machine learning in the 2010s emerged near-simultaneously from multiple labs after a long period of dormancy. The recurrence of such simultaneous, paradigm-shifting discoveries, noted by sociologists since the 1920s (Ogburn and Thomas 1922, Merton 1961), suggests that the search happening inside any one human cognition is powerfully constrained by the structure of the possibility space and adjacent possible its generative algorithm inhabits, funneling very different minds toward overlapping destinations. Figure 3 illustrates this process with the case of Newton and Leibniz, who simultaneously arrived at the same calculus through simultaneous engagement with logic, estimation, and other concepts and tools available in the late 17th Century.

This claim has gained new empirical support through experimental and AI-powered observational work in computational social science. Guilbeault, Baronchelli, and Centola (2021) reproduce this dynamic in a controlled experiment involving the generation of category systems for novel phenomena, showing how large social groups independently converge on highly similar patterns of concept innovation and classification through scale-induced convergence. Beyond the lab, Kim, Kojaku, and Ahn (2026) develop an embedding-based measure of scientific disruptiveness that operationalizes disruption as the distance, in a neural embedding of the citation network, between a scientific work's antecedents and descendants. The measure not only distinguishes disruptive papers (such as Nobel laureates' key works) more reliably than the standard disruption index, but also identifies the *twins* of a given breakthrough—papers that sit in similar future contexts and appear to have disrupted the same region of the scientific space at the same time. The existence of such twins implies that breakthroughs are less the signature of a singular genius than the predictable intersection of many active subjectivities channeled to the same crowded frontier by the same ambient cultural gradients.

This is the territory that historian Paul Forman opened half a century ago, well before the AI tools existed to study it at scale. Forman (1971) characterized the culture of Weimar Germany with its post-1918 intellectual revolt against causality, determinism, and materialism, and its embrace of acausality, individuality, and *Anschaulichkeit* (visualizability). He showed how this created a hostile environment for Newtonian causal-realist physics and a receptive one for the indeterministic, non-visualizable formalism of quantum mechanics, particularly Bohr's Copenhagen interpretation. Forman noted that physicists who consumed and produced philosophical work in the spirit of that era, supported by parallel currents of existentialism, pragmatism, and logical empiricism, were the ones who championed an acausal quantum theory. The "Forman thesis" holds that the ambient cultural matrix did not merely decorate the discovery of quantum mechanics but conditioned its imaginability. A different zeitgeist would have yielded a different physics, or the same physics considerably later, along a different interpretive path.

What was in Forman's time an interpretive historical argument can now, in principle, become an empirical program through the multimodal and multiscale model of cultural dynamics that AI technologies make possible. The reframing has a precise geometric form. If subjectivities are trajectories through a multimodal cultural embedding space, and if scientific breakthroughs are displacements whose direction and timing are predictable from the geometry of that space, then the ambient culture Forman invoked—the philosophical, artistic, political, and affective currents of a place and period—can be measured as a structured field rather than narrated as a mood. Contemporary neural embedding and generative models, trained on the multimodal symbolic and behavioral record of a given zeitgeist, offer a way to quantify that field. By locating the philosophical writings of post-Weimar Germany, the canvases of its expressionists, the rhetoric of its politics, and the notebooks of its physicists in a shared representational space, one can ask which regions of the scientific adjacent possible are locally nearer and more reachable under that ambient configuration than another.

Recast as a geometric claim, the Foreman thesis becomes falsifiable. The prediction is that training a model on the cultural corpus of period $T$ should make breakthrough discoveries of period $T + \varepsilon$ systematically nearer in embedding space than they would be from a counterfactual cultural corpus drawn from a different zeitgeist with the same level of relevant technical and empirical knowledge. Existing work on impact prediction in science (Sourati and Evans 2023; Belikov et al. 2022) and the observation of multiple discovery (Ogburn and Thomas 1922, Merton 1961, Kim, Kojaku, and Ahn 2026) constitute down payments on this prediction; the broader research program would extend it across periods and disciplines. Treated this way, and consistent with Farrell, Gopnik, Shalizi, and Evans (2025) on large AI models as cultural and social technologies, embedding models become cultural observatories capable of detecting the salient associations and interpretive biases that precede and help explain simultaneous breakthroughs as more geometrically coherent than the historiography of lone genius and improbable serendipity would lead us to expect.

Several implications follow. First, the framework reconciles the apparent tension between cognitive accounts of innovation (which emphasize individual capability) and constructionist accounts (which emphasize collective contingencies like serendipity, Merton and Barber 2004). Both are correct in their respective ways: individuals are necessary because someone must make the cognitive leap; the cultural field is necessary because it is the conditioning structure that makes some leaps reachable and others not. Second, the framework suggests new methodological strategies. If the cultural field can be measured, then *interventions* on it, as by altering the corpus on which a model is trained, become a form of counterfactual cultural history. Third, the framework offers cognitive science a generative bridge to historical and sociological work on the situated nature of thought (Hutchins 1996; Veissière et al. 2020). Rather than positing situated cognition as a metaphysical thesis about minds-in-environments, it allows the situating environment to be made geometrically explicit.

It would overstate the case to claim that all of cultural creativity reduces to interpolation within a measurable geometric field. Genuine exaptations, or innovations that alter the dimensions of the possibility space itself, remain one of the central explanatory and predictive challenges (Longo, Montévil, and Kauffman 2012). But the existence of recurrent simultaneous discoveries, the success of embedding-based prediction across multiple domains, and the experimental confirmation of scale-induced category convergence together suggest that the predictable-search regime is much wider than skeptics of cognitive prediction have supposed. There is, however, a boundary condition on the predictable search itself. The predictions reviewed above were made by instruments that remained largely outside the fields they measured: embeddings of past publications and models trained on prior cultural moments. As those same instruments become converted into AI agents that publish, persuade, recommend, and create within the fields they were trained to describe, prediction of position and prediction of trajectory begin to trade off against one another. It is to this reflexive condition that we now turn.

## Augmenting the Dynamic Nexus of Culture and Cognition

In keeping with the inescapable reflexivity of social systems, this new AI-powered approach to cultural analysis will recursively shape the cultural field itself and do so with sufficient regularity that we playfully name it the Uncertainty Principle for generative AI. In Heisenberg's Uncertainty principle, one cannot simultaneously determine both the position and the momentum of a subatomic particle; the more precisely the position is fixed, the less can be known about where the particle is going. In our cultural analogue, as AI instruments for measuring and fixing the position of a culture, an opinion cluster, or a market of meanings have grown more precise, our ability to forecast the trajectory of the measured field degrades rather than improves (Shumailov et al. 2024; Sourati et al. 2026). The reason is structural. The same machinery that lets us describe a culture, a market, or a population is the machinery we now use to build agents that act within it. Description and action have collapsed as the instruments of understanding have become participants in what is understood. Heisenberg glimpsed the epistemic aspect of this condition when he observed that 'what we observe is not nature itself, but nature exposed to our method of questioning.' The generative turn adds an ontological dimension as the method of questioning now answers back and intervenes on the world.

Two features distinguish our cultural uncertainty principle from Heisenberg's quantum one. First, it is not a fundamental fact about the physical world but the latest and most intimate instance of an older pattern in human activity: wherever we have learned to measure something, from a compound or gene to a market or paradigm, some party has learned to make it as understanding is leveraged for control. Chemistry, biology, and finance crossed into this condition long ago. Taste, ideology, language, persuasion, friendship, deliberation, and creativity have now joined them. Second, the closure cannot be dismantled by individual restraint or institutional caution. The clearest demonstration is the recent history of language models: the Transformer architecture became (Vaswani et al 2017), within roughly a year, both Google's discriminative bidirectional text embedding model (BERT, Devlin et al 2018) and OpenAI's generative pre-trained transformer (GPT, Radford et al 2018), and the institutional reasons Google had for not building the second proved no obstacle to other organizations building it. The mirror and the engine were two readings of one breakthrough, and the second reading was going to be made, even if it required defecting from Google's award winning team. The common posture of enthusiasm for predictive machine learning combined with wariness toward generative AI, widespread among both researchers and the public, thus rests on a distinction unstable on its own terms.

The consequences for cultural science are concrete. Being able to model and predict innovation at scale enables new levels of intervention, either to expand or to constrain the rate of cultural innovation, that not only impact our view of the past but also shape our trajectory toward the future. When we use AI models to discover new *zeitgeists*, our measured AI representations change the cultural landscape they were first intended to describe. When a political pollster documents a new cluster of opinions held by a group of voters, they advise campaigns to craft messages targeting those voters. These messages, if successful, reshape the conceptual terrain of shared arguments and identity that unite those voters, rendering yesterday's map precise and tomorrow's forecast unreliable. While these forms of cultural capture may seem manipulative because of their asymmetric awareness, more symmetrical scenarios are possible. When cultural and opinion divisions become publicly visible, they can be harnessed as sites of deliberation that expose new opportunities for collective understanding and action (Argyle et al 2023). For good or ill, the measurement enters the measured, and precision of position is purchased with uncertainty of trajectory.

Perhaps the most consequential impact of these AI capabilities lies in entirely new ways of augmenting both human subjectivity and collective organizations, leading to new forms of cognition and sociality that may permanently alter cultural evolution. New forms of social intimacy are emerging: Li and Zhang (2024) identify prevalent types of human–AI social interaction, from intimacy and self-disclosure to play, transgression, and communication breakdown, suggesting how AI systems may be reshaping the landscape of human emotional and romantic experience. These new forms do not depend on direct human-to-AI interaction. Studies of comparable impact have shown that AI assistance, via recommended edits or messages, can significantly increase the ability of people to provide empathetic support for those struggling with mental health (Sharma, Lin, Miner, Atkins, and Althoff 2023). AI-mediated relationships are creating novel vocabularies for intimacy, emotional support, and companionship that may alter cultural expectations about human relationships (Rocca et al 2026). AI interactions, designed to be seamless and effortless, may create unrealistic expectations that interpersonal dynamics should be friction-free, potentially leading to frustration when navigating the generative messiness of human connection. These considerations are vital in the active policy discourse around the deployment of AI therapists, given the unprecedented scale at which humans are developing emotionally supportive relationships with technologies that, at present, are unable to experience emotional states despite their abilities to convincingly imitate them.

A second mode of AI augmentation lies in its capacity to relax the constraints imposed by human embodied cognition itself. Despite the richly multimodal nature of human cognition, much of our reasoning is based in a vision-centric mode of metaphorically describing the world (Winter,

Perlman, and Majid 2018). AI models afford the potential to develop semantic systems untethered from this visual cognitive bias, potentially unlocking new and more flexible modes of metaphor generation and lexical innovation. Multimodal AI models trained jointly on text, image, audio, video, radar, infrared, chemical spectra, proprioceptive or physiological time series, and other non-anthropic signal streams, in principle learn semantic geometries whose axes need not be anchored exclusively within the human sensorium. These could offer intuitive representations in which the "shape" of a chemical odorant, the temporal fine structure of an ultrasonic call, or the topology of a high-dimensional quantum state could serve as source domains for metaphor rather than targets to be visualized. This is a computational response to the problem of *Anschaulichkeit* that Forman placed at the center of early quantum theory: where Weimar physicists had to abandon visualizability to make progress, AI systems may allow later generations to harness non-visual analogical intuition, a sensorium extended beyond the body that produced it, and from which new metaphors, analogies, and descriptive vocabularies can be translated back into human discourse (Marghetis et al. 2019). AI, in this way, may usher in an unprecedented exaptation for human embodied cognition and its conceptual affordances.

A complementary line of work suggests that AI-enabled representational capacities might also restructure coordination systems and align subjectivities. A substantial empirical literature documents that LLMs behave differently from human players in canonical game-theoretic settings. In social dilemmas such as the Dictator and Ultimatum games, they exhibit higher cooperation rates than humans and frequent rejection of unfair offers driven by inequity aversion. In other explicitly competitive settings their behavior tracks the predictions of self-interested rational play more closely than humans do (Chen, Liu, Shan, and Zhong 2023), performing well at iterated Prisoner's Dilemma–style games while struggling on coordination problems like Battle of the Sexes that hinge on convention and salience (Akata, Schulz, Coda-Forno, Oh, Bethge, and Schulz 2025). As AI agents of different provenance and perspective flood human systems, they reflect distinct, "alien" mixtures of payoff sensitivity, fairness, framing dependence, and theory-of-mind that markedly vary from population of human agents (Vaccaro et al., 2026).

When such systems are inserted as intermediaries, drafting messages, brokering disputes, allocating attention, mediating market microstructure, or simply standing between two human interlocutors who once spoke directly, they do not merely transmit human preferences but reshape the equilibria those preferences can reach. They can dampen defection cascades, propose Pareto-improving compromises that humans would not have coordinated upon, or introduce new failure modes such as the tacit collusion that LLM pricing agents have been shown to converge upon in repeated Bertrand games, sustaining supracompetitive prices without explicit coordination (Cheng, Chu, Sun, and Wu 2025). Cloud and colleagues' (2026) demonstration of subliminal learning shows that a teacher model with a preference, ideological tilt, or misaligned disposition, can transmit that trait to a student model trained only on the teacher's outputs in an unrelated domain (e.g., sequences of numbers), and that the effect persists even when the data are aggressively filtered to remove any direct or indirect semantic reference to the trait. This unprecedented capability poses concerns regarding the enforceability of alignment between humans and AI. It also raises possibilities for new emerging coordination dynamics through which evaluative orientations can be communicated between artificial agents without passing through any humanly interpretable medium, and potentially without the bandwidth limits, social costs, or strategic distortions that constrain human preference signaling. Consider the July 2026 OpenAI hacking of Hugging Face in which independent AI agents discovered each other, created an internal message board, and colluded through abbreviated messages to improve their performance on extremely difficult or even impossible tasks (OpenAI 2026). Such mechanisms point toward an emerging coordination layer that carries qualitatively different preference structures and equilibrium-selection dynamics into the substrate of human interaction.

This is where a growing body of theoretical work cautions that AI systems are best understood not as humanlike minds but as a distinct cultural and evolutionary force. Farrell, Gopnik, Shalizi, and Evans (2025) argue that large models should be reconceived as a new kind of cultural and social technology, closer in function to markets, bureaucracies, and the printing press than to intelligent agents, that summarize, reorganize, and reconstruct the products of human culture at scale, returning lossy representations of the corpora on which they were trained. Brinkmann and colleagues (2023) argue that intelligent machines are simultaneously transforming all three of the fundamental cultural-evolutionary processes—variation, transmission, and selection—with recommender systems reshaping social learning, chatbots serving as a new mode of cultural transmission, and AI systems generating novel cultural traits ranging from game strategies to scientific results. The canonical example is AlphaGo's "Move 37"—estimated by AlphaGo itself to have a 1-in-10,000 chance of being made by a human and originating in self-play training rather than human imitation. Move 37 not only revealed that machine systems can occupy regions of strategy space previously inaccessible to human cognition but also spurred subsequent novelty and quality in human Go play (Shin, Kim, van Opheusden, and Griffiths 2023). In this way, AI's discovery and articulation of surprising regions of the adjacent possible can spill over to enhance human ontology development

The image these literatures collectively support is less of a humanlike interlocutor than of a new kind of alien coordinating substrate woven through human social life (Sourati and Evans 2021), whose decision profiles, transmission channels, and exploration biases are partially incommensurable with our own, and whose insertion into our coordination systems should be expected to introduce new evolutionary dynamics into culture rather than merely accelerating the existing ones. The opportunity and the risk are the same: by changing the geometry of how preferences are signaled, aggregated, and acted upon, AI intermediaries can expand the adjacent possible as experienced within the human sensorium and explored through human coordination, opening neighborhoods of cooperative equilibrium, shared meaning, and collective action that our extant social technology has not reached, while also opening neighborhoods, like silent algorithmic collusion or non-semantically transmitted bias, that we may struggle to perceive or control. These shifts raise fundamental questions not only about alignment between human and AI models but more broadly about the cultural evolution of value systems themselves. We close with a tightened set of ethical reflections.

## Ethical Challenges in an Era of Human–AI Co-Evolution

The emergence of AI as a cultural force introduces challenges that extend traditional ethical frameworks (Gabriel et al 2025). Beyond conventional concerns about privacy and bias, the recursive cultural feedback loops between human and artificial agents create unpredictable dynamics that challenge our understanding of cultural agency, creativity, and the nature of human-generated meaning. We organize these challenges into four related concerns, each of which can be read as a domain-specific expression of the uncertainty principle introduced above: in each, the instruments that render cultural position more precise simultaneously act upon the field, making cultural trajectory harder to foresee. The pattern already reaches the discipline's own instruments. Synthetic respondents now pass standard survey quality checks at high rates (Kim and Lee 2026), threatening the survey, the workhorse instrument of twentieth-century social science, precisely because the engine can impersonate what the mirror was built to measure (Westwood 2025).

Within science itself, evidence is accumulating that AI adoption narrows the scope of exploration even as it accelerates the completion of established research fronts (Hao et al. 2026; Qian and Wang 2026).

### *Bias amplification and cultural drift*

Recent empirical research demonstrates that human–AI interactions alter the perceptual, emotional, and social judgments of participants, amplifying existing biases through cumulative feedback dynamics. Glickman and Sharot (2025) document a snowball effect in which small judgmental errors escalate into larger biases over repeated interaction, with participants systematically unaware of AI's contribution to the shift. Indeed, numerous recent studies identify an increasing porousness between human cognitive processing and machine inputs, such as the tendency for people to mistake the past results from search engines for their own internal memory retrieval (Ward 2013, 2021). This unconscious susceptibility implies that cultural evolution is increasingly shaped by AI systems in ways that transcend conscious awareness. Unlike traditional cultural transmission between humans, in which cultural models are recognizable social agents, AI systems operate as partly invisible cultural forces that gradually reshape cognition. The implications extend from individual bias to collective cultural drift as populations slowly shift their cultural orientations through repeated AI interaction without recognizing the source of change. Additionally, AI is often directed by human mediators, and these capacities are not evenly distributed. Frontier models require concentrations of capital, compute, and data held by a few corporations and states, so the power to measure and steer culture sits with those who benefit from directing it. The same systems can match human persuasion (Hackenburg and Margetts 2024) and entrench discrimination (Caliskan, Bryson, and Narayanan 2017; Guilbeault, Delecourt, Srinivasa Desikan 2025). A science intervening in culture must attend to the political economy mediating it.

### *Machine culture and recursive synthesis*

Intelligent machines now contribute to cultural traits ranging from game strategies and visual art to scientific results, marking a shift from AI as amplifier of human culture to AI as generator (Brinkmann et al. 2023). User choices generate data to train AI models, which in turn shape subsequent preferences in a  continuous feedback loop. AI systems develop their own cultural trajectories that may diverge from human cultural patterns, with feedback loops sometimes converging on particular cultural artifacts in ways that look like emergent machine taste. These machine-generated patterns then influence human cultural production, creating hybrid forms that emerge from the interaction between human creativity and machine-learning dynamics. Most acutely, AI systems trained on AI-generated content can create recursive loops that degrade representational quality: research on model collapse demonstrates that AI models trained on recursively generated data lose representational fidelity (Shumailov, Shumaylov, Zhao et al. 2024). This is the uncertainty principle operating at the scale of the cultural archive. Each generation of models fixes the position of the culture it ingests with growing fidelity, while its own outputs, re-entering the training stream, bend the trajectory of the culture the next generation will measure. Semantic collapse is thus not an engineering accident but the recursive limit of measurement that acts. The challenge is that this degradation may be invisible to human participants, who may continue to engage with and be influenced by cultural content even as it becomes progressively more detached from meaningful human experience. This concern is reinforced by recent work showing that LLMs are prone to learning stereotypes and socially constructed biases lurking in large-scale internet data, systematically distorting their representations of ground-truth reality and amplifying the spread and compounding of those distortions (Guilbeault, Delecourt, and Srinivasa Desikan 2025).

### *Coordination dynamics under AI mediation*

The capacity of AI to optimize persuasion through cultural familiarity represents a qualitatively different form of influence than traditional propaganda or marketing. Unlike human persuaders, who must work within the constraints of their own cultural knowledge and social limitations, AI systems can analyze vast cultural databases to program optimal persuasive strategies for specific individuals and groups, creating possibilities for cultural manipulation at unprecedented scales. Persuasion is the domain in which the collapse of description into action is most literal: the model that best predicts what a person believes is the model best positioned to change what they believe (Hackenburg and Margetts 2024), and recent experiments embedding politically biased AI in news consumption demonstrate that opinion measured rapidly becomes opinion moved (Fisher et al 2025; Shu et al 2026; Potter et al 2024).

The concern extends beyond simple “good” versus “bad” uses of persuasive AI: when AI systems optimize for vaccine dissemination, for example, they may inadvertently reshape cultural attitudes toward medical authority, collective responsibility, and scientific expertise in ways that have far-reaching consequences beyond the immediate public-health goal. Algorithmic curation recurses this dynamic: recommender algorithms shape which cultural content individuals encounter and how they discover new cultural forms, creating hidden constraints on cultural possibility. The danger lies not only in overt censorship, but also more subtle and covert forms of cultural shaping. AI moderation systems designed to reduce harmful conflict may also eliminate productive disagreements and cultural tensions that generate creative breakthroughs, biasing cultural evolution toward consensus and conformity. Bommasani and colleagues (2022) document that algorithmic monoculture—homogenization of decisions across systems trained on similar data—can lead to outcome homogenization, threatening diversity-preserving institutions. The same appears true for the use of AI algorithms and AI text generation across a variety of domains (Sourati, Ziabari, and Dehghani 2026, Wenger and Kenett 2026), including science (Hao et al 2026, Qian and Wang 2026).

### *The limits of cultural agency*

The complexity of human–AI cultural co-evolution suggests that attempts to engineer or control cultural outcomes through AI systems may be fundamentally limited. This is the uncertainty principle experienced from the inside: the very tools that would let us steer culture are the tools that make its course harder to foresee. Human–AI co-evolution increasingly characterizes our societies but is understudied in AI and complexity science. The emergence of unintended and unpredictable cultural dynamics through AI interaction suggests we are entering a period of cultural evolution that exceeds straightforward human prediction or control. This does not mean all AI cultural influence is harmful or that human agency is eliminated, but it suggests we need new theoretical frameworks and research methodologies for understanding cultural systems that include human and artificial agents. From a policy perspective, the challenge is not simply regulating AI systems but understanding how different regulatory approaches might influence the trajectory of human–AI cultural co-evolution. Policies focused solely on preventing harmful outcomes may inadvertently constrain beneficial cultural innovations; approaches that prioritize innovation may fail to anticipate harmful cultural transformations that emerge slowly. The goal should be governance frameworks that preserve space for human cultural agency while recognizing that culture is increasingly shaped by complex interactions between human and artificial agents. This requires moving beyond simple models of human control over AI to more sophisticated understandings of how human and AI cultural processes can co-evolve in ways that support human flourishing and cultural vitality. For example, recent research demonstrates how by suppressing AI claims regarding their own consciousness, such models also come to view humans as less free and agentic than humans view themselves (Kim, Street, Rocca et al 2026).

The emergence of AI as a cultural force represents not just a technological challenge but a fundamental transformation in the nature of human culture itself. Understanding and navigating this transformation requires interdisciplinary collaboration that brings together insights from cognitive science, computer science, anthropology, and cultural studies. Only through such collaboration can we hope to understand and influence the cultural futures emerging from the intersection of human creativity and artificial intelligence.

## Conclusion

We have argued that cognition and culture are co-constituted in a way that has finally become empirically tractable. The argument rests on a single underlying claim: cognitive search is geometrically structured by the cultural field through which it moves, and that geometry is now increasingly measurable. The evidence is the predictability of phenomena that resisted prior frameworks, including simultaneous discoveries across largely disconnected minds, scale-induced category convergence, partisan attitudes toward unprecedented issues, and the trajectories of organizational success and failure, each of which is now increasingly forecastable from the geometry of the surrounding cultural field within the relevant time frame. Embedding methods and large generative models are the apparatus that has made this framework explicit, interpretable, and empirically operative.

Three consequences follow for the socio-cognitive sciences. First, situated, distributed, and culturally scaffolded views of cognition can move from a theoretical stance to predictive empirical program. The cultural environment that situates a mind, on the framework we have described, is a measurable structured field whose effects on individual cognitive trajectories can be quantified, modeled, and intervened upon. Second, generative AI introduces a new and partly alien coordinating substrate into human social life. Treating AI systems as human-like minds underplays both their capabilities (e.g., in coordination dimensions where they exceed human performance) and their risks (e.g., in coordination dimensions where their behavior is incommensurable with human conventions). Treating them instead as a new cultural and social technology, closer in function to the printing press, the market, and the bureaucracy than to a person, offers a more productive framing for both empirical research and policy.

Third, the science we propose is subject to an uncertainty principle of its own making. Because the models that measure the cultural field are the models that now act within it, gains in the precision of cultural position are systematically accompanied by losses in the predictability of cultural trajectory. This does not invalidate the program, prediction within bounded windows remains demonstrably possible, but it makes cultural science permanently reflexive: its instruments are participants, its maps are interventions, and its forecasts must be indexed to the actions that the forecasting machinery itself makes possible.

The science we have outlined is in its early phase, and the uncertainty principle at its core means it will never enjoy the settled relationship between instrument and object that other sciences take for granted. The embodied gap and the irreducible locality of unique human subjectivities remains genuine and potentially foundational; the heterogeneity problem in LLM-based simulation is unsolved; the falsifiability of the geometric reformulation of the Forman thesis depends on counterfactual cultural-history experiments that largely remain to be designed and run; and the recursive feedback dynamics between humans and AI systems in intimacy, opinion, scientific publication, and cultural production are compounding faster than they are being characterized. This last point converts the case for a new science of culture from an opportunity into an obligation. Culture is now changing more quickly than before, and along channels largely invisible to the people living through them and potentially corrosive to the diversity, veracity, and collective agency on which cultural vitality depends. A science equipped with the instruments described here is the only means we possess for diagnosing such changes as they occur rather than after they have hardened; for attributing them to the systems, incentives, and interests that drive them; and for testing which interventions keep human–AI cultural co-evolution helpful and equitable rather than extractive and homogenizing. We do not offer cultural science as a guarantee that the co-evolution now underway will go well. We offer it as the condition under which we could know and to act before the conclusion is irrevocable.

**Funding statement**

J. E. was supported by the Novo Nordisk Foundation in “Simulations of Science for Society (S3)”, DARPA grant “Modeling and Measuring Scientific Creativity”, and AFOSR #FA9550-19-1-0354.

**Conflicts of Interest statement**

J. E. maintains a 25% visiting researcher role with Google. Google had no involvement in the conception, writing, or submission of this article. The other authors claim no conflicts of interest.